\documentclass[twocolumn]{aastex631}
\usepackage{amsmath,amssymb,booktabs,array}
\usepackage{xcolor}

\newcommand{\pauline}[1]{#1}

\graphicspath{{figures/}}

\newcommand{\Msun}{M_\odot}
\newcommand{\mbh}{M_\bullet}
\newcommand{\LR}{L_{\rm R}}
\newcommand{\LX}{L_{\rm X}}
\newcommand{\LEdd}{L_{\rm Edd}}
\newcommand{\mdot}{\dot m}
\newcommand{\ellx}{\ell_{\rm X}}
\newcommand{\lamx}{\lambda_{\rm X}}
\newcommand{\xirx}{\xi_{RX}}
\newcommand{\muFP}{\mu_{\rm FP}}
\newcommand{\xij}{\xi_j}
\newcommand{\DeltaR}{\Delta_{\rm R}}

\shorttitle{SKA/ngVLA Inverse-FP IMBH Searches}
\shortauthors{Abbassi et al.}

\begin{document}

\title{Validating Inverse Fundamental-Plane IMBH Mass Estimates in the SKA/ngVLA Era}

\author[0000-0003-0428-2140]{Shahram Abbassi}
\affiliation{Department of Physics \& Astronomy, University of Western Ontario, London, Ontario, Canada}
\email{sabbassi@uwo.ca}
\author[0000-0002-5730-0376]{Samik Mitra}
\affiliation{International Centre for Theoretical Sciences, Tata Institute of Fundamental Research, Bengaluru, Karnataka 560089, India}
\email{Corresponding Author: samik.mitra@icts.res.in}

\author[0009-0002-0373-570X]{Ramananda Santra}
\affiliation{International Centre for Theoretical Sciences, Tata Institute of Fundamental Research, Bengaluru, Karnataka 560089, India}

\begin{abstract}
The radio/X-ray/black-hole-mass Fundamental Plane (FP) is widely used to estimate masses of accreting intermediate-mass black hole (IMBH) candidates. Its inverse use is physically justified only when the radio luminosity traces a compact jet core and the X-rays arise from the same sub-Eddington hard-state accretion flow. We formulate an accretion-state validity criterion for inverse-FP masses using verified black-hole X-ray binary (BHXB) radio/X-ray tracks as the empirical reference. {{Using 424 detections from 21 BHXBs and candidates, including recent radio/X-ray monitoring, we find that both standard and curved tracks are present and that the luminosity at which the radio/X-ray behavior changes is source dependent. A low X-ray Eddington ratio is therefore not, by itself, sufficient to establish inverse-FP validity.}} {{Within a scale-invariant jet/hot-flow framework, track steepening reduces the effective mass leverage of the inversion, while at much lower accretion rates uncertainties in the dominant emission mechanism and compact-core production can invalidate the underlying FP assumptions altogether.}} We apply the criterion to representative IMBH-relevant systems without fitting a new plane. NGC 4395 is low-Eddington but compact-radio deficient: its VLA-scale and VLBI jet-base luminosities imply formal inverse-FP mass underestimates of about 1.0 and more than 2.8 dex if the deficit is misread as a mass signal. Globular-cluster X-ray populations in NGC 1399 and NGC 4472 are source-identity limited: 71\% and 67\% lie below $L_{\rm Edd}(10\,M_\odot)$, and the high-luminosity tail alone does not identify an IMBH accretor. The radio non-detection of the $\omega$ Centauri IMBH candidate constrains gas supply, accretion efficiency, and compact-core production rather than excluding the black hole. The resulting test defines which SKA/ngVLA-era radio cores can be converted into FP masses, and which instead constrain gas supply, source identity, or compact-core production.
\end{abstract}



\section{Introduction}\label{sec:intro}

Intermediate-mass black holes (IMBHs) occupy the mass range between stellar remnants and supermassive black holes, but they remain the least secure part of the black-hole demographic census. Their expected spheres of influence are small, their accretion luminosities are often weak, and many candidate environments---dwarf nuclei, dense star clusters, and off-nuclear X-ray sources---contain stellar binaries, supernova remnants, background AGNs, and star-forming regions that can mimic or contaminate accretion signatures. Establishing whether a source is an IMBH therefore requires not only sensitivity, but also a physically controlled mass diagnostic.

The radio/X-ray/black-hole-mass Fundamental Plane (FP) provides one of the few observational tools that can connect stellar-mass black holes and low-luminosity AGNs in a common accretion framework \citep{Merloni2003,Falcke2004,Gultekin2009,Plotkin2012,Gultekin2019}. Its physical motivation is the scale invariance of radiatively inefficient accretion flows and compact, self-absorbed jets \citep{HeinzSunyaev2003,YuanNarayan2014}. Because radio and X-ray observations are feasible even when dynamical measurements are not, the FP has naturally become a major ingredient in IMBH searches. It underlies early radio tests for black holes in globular clusters and dwarf spheroidals \citep{Maccarone2004,MaccaroneFenderTzioumis2005}, ngVLA-oriented searches in Virgo globular clusters \citep{Wrobel2021}, and SKA forecasts for the NGC~1399 globular-cluster system \citep{Karimi2024}. This issue becomes especially important for SKA-era IMBH programs, where the increased supply of faint radio detections and upper limits will be scientifically useful only when compact, positionally coincident jet-core emission can be separated from diffuse or background radio components.

The difficulty is that the FP is often applied in reverse. Given a radio luminosity and an X-ray luminosity, one solves for a black-hole mass. This procedure can be useful, but it implicitly assumes that the measured radio flux is the compact jet core and that the X-ray luminosity belongs to the same sub-Eddington hard-state accretion regime that generates the plane. These assumptions are not automatic. In this context, a radio detection is not automatically an FP-usable radio luminosity: only a compact, positionally coincident jet-core component can be interpreted as the radio term in the plane. The same candidate can be assigned different masses depending on the adopted FP calibration, non-detections are sometimes interpreted as excluding dynamically motivated IMBHs, and luminous globular-cluster X-ray sources can be treated as IMBH candidates even when stellar-remnant binaries are sufficient. Recent low-mass AGN studies also show that FP behavior depends on Eddington ratio, radio loudness, and source classification \citep{Bariuan2022,Gultekin2022,Wang2024}. The inconsistency is therefore not only a question of regression coefficients; it is a question of whether the source satisfies the physical domain of the FP.

{{The physical limitations of an inverse-FP estimate can arise in more than one accretion regime. At higher hard-state luminosities, changes in radiative efficiency and in the radio/X-ray relation can weaken the mass dependence on which the inversion relies. At much lower Eddington ratios, the problem is different: a compact self-absorbed jet, the origin of the X-ray emission, and the relevant radiative efficiency cannot be assumed a priori. Gas starvation, strong outflows, inefficient particle acceleration, or a change in the dominant emission component may therefore invalidate an FP mass estimate even when the source lies well below the transition regime \citep{YuanNarayan2014,Plotkin2012}. The X-ray Eddington ratio is consequently useful as a state indicator, but not as a stand-alone validity test for the inverse FP.}}

This paper develops a conservative accretion-state criterion for inverse-FP IMBH masses. The analysis is deliberately separated into three levels. First, we measure the relevant state behavior in verified BHXB radio/X-ray tracks, where accretion states are best sampled. Second, we use scale-invariant jet/hot-flow physics to connect a steepening radio/X-ray track to a loss of effective mass leverage in an inverse FP. Third, we apply the resulting criterion to IMBH-relevant systems only as a diagnostic, not as a new universal calibration. The goal is to decide when the FP is allowed to constrain an IMBH mass, and when it should instead be treated as a constraint on source identity, compact-core production, or gas supply.

Throughout, $\LR\equiv\nu L_\nu$ at 5~GHz and $\LX$ denotes X-ray luminosity. Radio measurements reported at frequencies other than 5~GHz are converted using $S_\nu\propto\nu^\alpha$, adopting the measured spectral index where available and a flat compact-core convention otherwise; the assumed spectral index is treated as a radio systematic, especially for weak or non-simultaneous measurements. We distinguish the dimensionless X-ray Eddington ratio $\ellx\equiv\LX/\LEdd$ from its logarithmic state coordinate $\lamx\equiv\log_{10}(\LX/\LEdd)$. \pauline{Here $\lamx$ is used only as a dimensionless logarithmic accretion-state coordinate; it is not a wavelength.} For the BHXB calibration, we use the 1--10~keV luminosities provided by the Bahramian et al. database. The notation and FP reference relation are defined in Section~\ref{sec:physics}.


\section{Physical criterion}\label{sec:physics}

We use the canonical FP as a diagnostic reference,
\begin{equation}
 \log \LR = \xirx\log\LX + \muFP\log\!\left(\frac{\mbh}{\Msun}\right) + b_0,
 \label{eq:fp}
\end{equation}
with $\xirx\simeq0.60$, $\muFP\simeq0.78$, and $b_0\simeq7.33$ in cgs units for the original mixed XRB/AGN calibration \citep{Merloni2003}. In this paper these coefficients define a reference ridge; they are not assumed to be valid for every IMBH candidate.

For a scale-invariant, partially self-absorbed compact jet, the core radio luminosity scales as \citep{HeinzSunyaev2003}
\begin{equation}
 \LR \propto \mbh^{\xij}\mdot^{\xij}, \qquad
 \xij = \frac{17}{12} \approx 1.42,
 \label{eq:jet}
\end{equation}
for a nearly flat radio spectrum. For a hot, radiatively inefficient X-ray flow \citep{YuanNarayan2014},
\begin{equation}
 \LX \propto \mbh\,\mdot^q,
 \label{eq:xray}
\end{equation}
where $q\simeq2$--$2.3$ in the deep hard state and approaches unity as the flow becomes radiatively efficient. Eliminating $\mdot$ between Equations~(\ref{eq:jet}) and (\ref{eq:xray}) gives
\begin{equation}
 \xirx = \frac{\xij}{q}, \qquad \muFP = \xij\!\left(1 - \frac{1}{q}\right).
 \label{eq:budget}
\end{equation}
Equation~(\ref{eq:budget}) implies $\muFP=\xij-\xij/q=\xij(1-1/q)$, which equals $\xij-\xirx$ only because $\xirx=\xij/q$. This algebra holds exactly in the idealized single-zone ADAF/compact-jet model; departures from either scaling law will break the identity. The inverse-FP mass leverage is therefore not an arbitrary regression coefficient; {{within this limiting model, it is the exponent of the residual mass dependence after $\mdot$ has been eliminated, so that $\LR\propto\LX^{\xirx}\mbh^{\muFP}$ with $\muFP=\xij-\xirx$.}}

Equation~(\ref{eq:budget}) is the useful physics, but it must not be over-interpreted. Near the hard-to-soft transition the assumptions behind Equations~(\ref{eq:jet})--(\ref{eq:xray}) begin to fail: the compact jet can weaken or disappear, the X-ray band can become disk/corona dominated, and hysteresis makes a single luminosity threshold imperfect \citep{FenderBelloniGallo2004,Belloni2010,Coriat2011,Corbel2013}. In that regime it is more accurate to state that the inverse FP loses validity than to claim that a universal value of $\muFP$ has been measured. Mathematically, the propagated mass uncertainty scales as
\begin{equation}
 \sigma^2_{\log \mbh} \simeq
 \frac{\sigma^2_{\log\LR} + \xirx^2\,\sigma^2_{\log\LX}
       + \sigma^2_{\rm int}}{\muFP^2},
 \label{eq:uncertainty}
\end{equation}
so an inversion becomes ill-conditioned as $\muFP\to0$ or whenever $\LR$ is not the compact-jet luminosity. Equation~(\ref{eq:uncertainty}) is a first-order propagation; it does not capture covariance between $\LR$ and $\LX$ or the non-Gaussian tails that arise from log-normal scatter in $\xirx$. It is presented as an order-of-magnitude guide to the ill-conditioning, not a precision error budget.

{{The applicability of the inversion also depends on whether the observed radio and X-ray components obey the assumed jet/hot-flow scalings. This remains true even well below the transition regime: an absent or unusually weak compact radio core reflects a breakdown of the model assumptions rather than, by itself, a vanishing mass coefficient.}}

This leads to a practical criterion. An inverse-FP mass is physically defensible only if the source passes four tests:
\begin{enumerate}
 \item \textit{Accretor identity:} the compact-object class is established independently of the FP.
 \item {\textit{Accretion state:} {the source should have independent spectral or timing evidence for an accretion regime compatible with the FP calibration. In the BHXB sample considered here, $\lamx\lesssim-2.5$ is useful only as an approximate screening scale in the 1--10~keV band. It is not treated as a fitted transition luminosity, and the source-dependent track changes discussed in Section~\ref{sec:additional} show that low $\lamx$ alone does not establish the required radiative efficiency, X-ray origin, or compact-jet behavior.}}
 \item \textit{Compact-core radio:} the radio emission must be associated with the unresolved candidate accretor rather than star formation, a supernova remnant, extended jet/lobe emission, or a background source. A spectrally flat or mildly inverted core, with $\alpha\gtrsim-0.1$ over 5--10~GHz where $S_\nu\propto\nu^\alpha$, provides strong evidence for a self-absorbed compact jet. Compact here means that the projected emitting region is constrained to be of order $\lesssim1$--$10$~pc for nearby low-mass AGNs. A brightness-temperature lower limit $T_B\gtrsim10^7$~K provides strong evidence for a compact non-thermal jet base, but a lower formal limit should not by itself reject a weak unresolved source if the angular resolution is insufficient. For Galactic and globular-cluster targets, the corresponding requirement is a stellar-position-coincident unresolved radio source, rather than cluster-scale diffuse emission, a supernova remnant, or a background AGN.
 \item \textit{FP residual check:} the canonical FP residual
       \begin{equation}
        \begin{aligned}
        \DeltaR \equiv {}& \log\LR - \bigl[0.60\log\LX\\
          &{}+0.78\log(\mbh/\Msun)+7.33\bigr]
        \end{aligned}
        \label{eq:residual}
       \end{equation}
       is not a large, coherent radio deficit or excess. Evaluating $\DeltaR$ requires an independently known $\mbh$; it is therefore a consistency check for sources with prior mass estimates, not an independent criterion for unknown masses.
\end{enumerate}
The first three conditions are physical prerequisites; the residual is a sanity check. Different failures bias the inverse FP in different directions: extended jet/lobe contamination, star formation, supernova-remnant contamination, and background association generally bias $\LR$ high and can overestimate $\mbh$, whereas compact-core suppression or jet quenching biases $\LR$ low and can underestimate $\mbh$ or produce misleading upper limits. Non-simultaneity can bias the residual in either direction.


\section{Data and diagnostics}\label{sec:data}

{{The BHXB diagnostic comprises 424 radio/X-ray detections from 21 physical systems, including BH candidates. It combines 269 measurements from the \citet{Bahramian2023} archival database (release v220908, 2022) with 39 detections from \citet{Carotenuto2021}, 32 from \citet{Hughes2025}, 16 from \citet{Shaw2021}, and 68 direct MeerKAT/Swift pairs from \citet{CrookMansour2026}. All radio luminosities are expressed as 5~GHz $\nu L_\nu$ and X-ray luminosities in 1--10~keV. Section~\ref{sec:additional} describes the selection, frequency conversion, and campaign-overlap criteria. The archival luminosity table does not provide individual observing dates or the distances used to derive each luminosity, so its quasi-simultaneity and distance scale are inherited from the published database rather than independently reconstructed here. Upper limits are retained in the data-selection ledger but are not included in the slope fits. The resulting fits are therefore conditional on the detected pairs and do not account explicitly for censoring or correlated variability.}}

The IMBH-scale objects and globular-cluster populations are used only as applications of the diagnostic; they do not enter the BHXB slope fits. This separation is important: the BHXB entries support the quantitative state measurement, whereas the IMBH and globular-cluster rows are published case studies used to identify which prerequisite of inverse-FP mass estimation fails.

\begin{figure*}[tbp]
\centering
\includegraphics[width=0.98\textwidth]{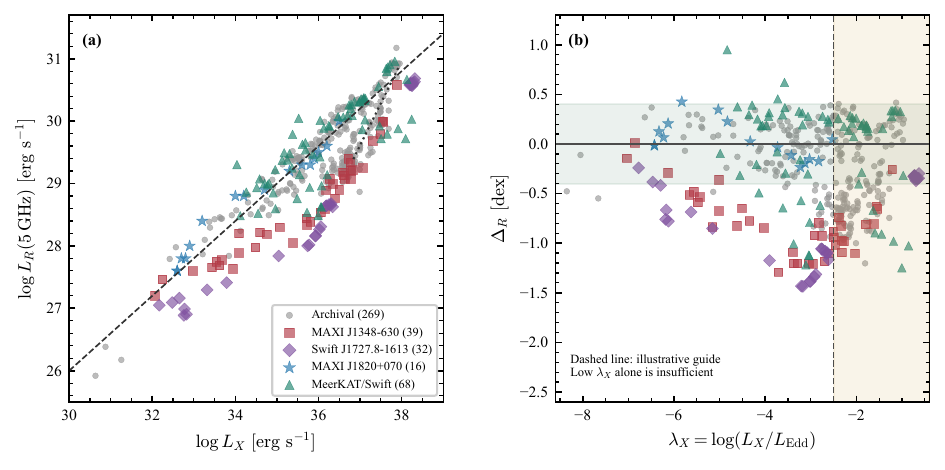}
\caption{{{BHXB radio/X-ray diagnostic for 424 detection pairs from 21 physical systems. Colors and symbols identify the contributing data sets; the MeerKAT/Swift subset includes BH candidates. Panel (a) shows 5~GHz $\nu L_\nu$ against 1--10~keV luminosity. Dashed and dotted guides illustrate slopes of 0.60 and 1.35, respectively; the dotted guide is normalized to the median fixed-slope intercept of the higher-$\lamx$ half of archival H~1743$-$322 at $\log\LX=37$ and is drawn only across the luminosity range sampled by that subset. Panel (b) shows the corresponding reference-FP residuals. The $\pm0.4$ dex band and $\lamx=-2.5$ line are included as visual reference scales rather than fitted boundaries. Absolute positions depend on the adopted distances and masses, and the figure does not imply a universal radio-quiet normalization or temporal ordering of the measurements.}}}
\label{fig:xrbtracks}
\end{figure*}

The radio-loud/radio-quiet classification adopted here follows \citet{Gallo2012} and \citet{Coriat2011}, who assigned track membership based on the radio luminosity offset from the canonical GX~339$-$4 correlation at matched $\LX$. Here ``radio-loud'' and ``radio-quiet'' refer to the BHXB track classification relative to the canonical GX~339$-$4-like radio/X-ray relation, not to the classical AGN radio-loudness parameter. {{Figure~\ref{fig:xrbtracks} shows the contributing data sets in the luminosity and residual planes. Systems without a secure historical track assignment are left unclassified, while the within-source curvature diagnostic is introduced below. These labels describe the observed track behavior used in the literature and are not intended to imply immutable source classes.}}

\begin{table*}[tbp]
\centering
\scriptsize
\setlength{\tabcolsep}{2.5pt}
\renewcommand{\arraystretch}{1.08}
\caption{Astrophysical samples, measured quantities, and permitted inference. Only the BHXB epochs enter the fitted radio/X-ray state diagnostic; all IMBH and globular-cluster entries are external applications used to test source identity, compact-core validity, or non-detection interpretation.\label{tab:provenance}}
\begin{tabular*}{\textwidth}{@{\extracolsep{\fill}}llll@{}}
\toprule
\parbox[t]{0.16\textwidth}{Sample} &
\parbox[t]{0.34\textwidth}{Data used} &
\parbox[t]{0.25\textwidth}{Provenance} &
\parbox[t]{0.20\textwidth}{Role in this paper} \\
\midrule
\parbox[t]{0.16\textwidth}{BHXB radio/X-ray tracks} &
\parbox[t]{0.34\textwidth}{{{424 detection pairs from 21 BHXBs/candidates;}} $\LR$ at 5~GHz; $\LX$ in 1--10~keV. Only detections are used in the fitted slopes; {{upper limits are retained for data provenance but not included in the regression.}}} &
\parbox[t]{0.25\textwidth}{{{\citet{Bahramian2023}, \citet{Carotenuto2021}, \citet{Hughes2025}, \citet{Shaw2021}, and \citet{CrookMansour2026}.}}} &
\parbox[t]{0.20\textwidth}{Quantitative state diagnostic; no IMBH candidate enters the fit.} \\[3pt]
\parbox[t]{0.16\textwidth}{NGC~4395} &
\parbox[t]{0.34\textwidth}{Reverberation mass, $\log(\mbh/\Msun)=5.56$; $\LX$; VLA 5~GHz luminosity; VLBI jet-base upper limit.} &
\parbox[t]{0.25\textwidth}{\citet{Peterson2005}; \citet{King2013}; \citet{Wrobel2006}; \citet{Yang2022}.} &
\parbox[t]{0.20\textwidth}{Compact-core audit; not an FP calibration anchor.} \\[3pt]
\parbox[t]{0.16\textwidth}{NGC~1399 globular clusters} &
\parbox[t]{0.34\textwidth}{77 GC-associated Chandra sources in 0.5--8~keV; bolometric correction $\kappa=2.0$ used only for Eddington-reference lines; previous SKA/FP context.} &
\parbox[t]{0.25\textwidth}{\citet{Lehmer2020}; \citet{Karimi2024}.} &
\parbox[t]{0.20\textwidth}{Population-level identity test; luminosities do not require IMBH accretors.} \\[3pt]
\parbox[t]{0.16\textwidth}{NGC~4472 globular clusters} &
\parbox[t]{0.34\textwidth}{30 GC X-ray sources; Virgo GC radio-search context.} &
\parbox[t]{0.25\textwidth}{\citet{Maccarone2003}; \citet{Maccarone2004}; \citet{Wrobel2021}.} &
\parbox[t]{0.20\textwidth}{Independent GC comparison; survey-selection context.} \\[3pt]
\parbox[t]{0.16\textwidth}{$\omega$~Centauri} &
\parbox[t]{0.34\textwidth}{Dynamical IMBH constraint; ultradeep radio upper limit.} &
\parbox[t]{0.25\textwidth}{\citet{Haberle2024}; \citet{Mahida2026}.} &
\parbox[t]{0.20\textwidth}{Non-detection as a constraint on gas supply and compact-core production.} \\
\bottomrule
\end{tabular*}
\vspace{2pt}
\begin{minipage}{0.98\textwidth}
\footnotesize\textit{Note.} The table is a provenance and usage map, not a homogeneous calibration catalog. The BHXB row supplies the only data used to fit or summarize the state-resolved radio/X-ray behavior. The other rows are published applications used to evaluate which physical prerequisite of inverse-FP mass estimation fails: source identity, compact-core validity, or non-detection interpretation. The GC source bandpass (0.5--8~keV) differs from the BHXB calibration band (1--10~keV); the Eddington reference lines in Figure~\ref{fig:gcs} are therefore illustrative identity checks, not BHXB-calibrated state boundaries.
\end{minipage}
\end{table*}

{{Figure~\ref{fig:xrbtracks} presents the BHXB sample in the radio/X-ray luminosity and reference-FP residual planes. Allowing each source or campaign its own intercept, the historically radio-loud series have a common descriptive slope $\xirx=0.575\pm0.010$ (conditional OLS uncertainty), compared with $0.600\pm0.013$ for the archival Bahramian subset. The close agreement with the canonical hard-state slope is accompanied by measurable source-to-source curvature: the independently fitted MeerKAT GX~339$-$4 campaign, for example, shows a positive median-split slope change. The radio-quiet and hybrid tracks likewise span a range of curvature and radio deficits, including behavior below $\lamx=-2.5$. The value $\lamx=-2.5$ is therefore shown only as a reference scale and is not used as a fitted population boundary or as a criterion for FP applicability.}}

\begin{figure}[tbp]
\centering
\includegraphics[width=\columnwidth]{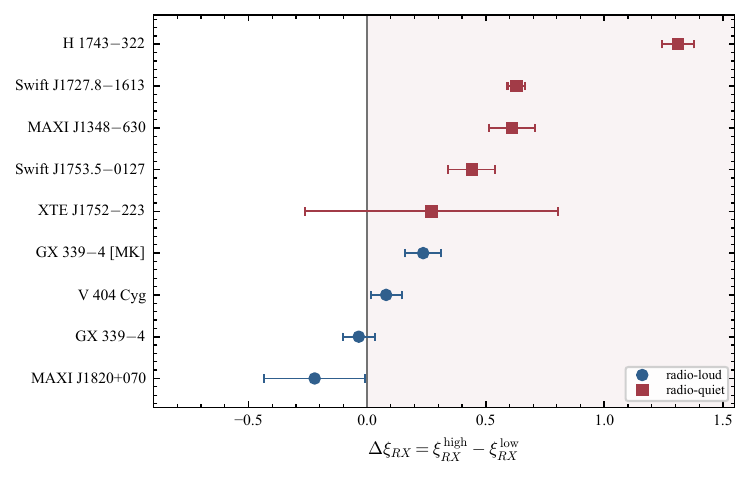}
\caption{{{Median-split curvature diagnostic for nine sufficiently sampled source/campaign series; XTE~J1118$+$480 is omitted because its higher-$\lamx$ fit is poorly constrained (Table~\ref{tab:slopes}). Radio-loud and radio-quiet denote the historical track classifications. MAXI~J1348$-$630 and Swift~J1727.8$-$1613 show clear positive curvature, MAXI~J1820$+$070 is consistent with no steepening, and the independent MeerKAT GX~339$-$4 campaign also shows a positive slope change. The light red region denotes $\Delta\xirx>0$. Error bars represent the conditional bootstrap uncertainties of the sampled detection pairs.}}}
\label{fig:stateslopes}
\end{figure}

The state-resolved slope change is summarized in Figure~\ref{fig:stateslopes} and Table~\ref{tab:slopes}. {{For each sufficiently sampled source or campaign, the detections are divided at the median $\lamx$, with values strictly below the median assigned to the lower subset and the remainder to the higher subset. Each half is resampled 2500 times, an ordinary least-squares relation is fitted in logarithmic luminosities, and the median slope and bootstrap standard deviation are reported.}} This simple diagnostic is not a substitute for detailed outburst modelling, but it addresses a physically direct question: {{whether the radio/X-ray relation changes systematically across the luminosity range sampled by an individual source.}}

{{The uncertainties reported in Table~\ref{tab:slopes} describe the conditional bootstrap dispersion of these within-source fits rather than the full uncertainty of an FP mass estimate.}} They do not include common-mode systematics from source distance, black-hole mass, radio spectral conversion, or X-ray bandpass conversion. This distinction is important because the present diagnostic uses within-source curvature, for which common-mode shifts largely translate tracks without creating a false steepening. The slopes are therefore used only to identify state-dependent track curvature; the paper does not use them as a new universal FP calibration.

\begin{table*}[tbp]
\centering
\scriptsize
\setlength{\tabcolsep}{4pt}
\renewcommand{\arraystretch}{1.12}
\caption{State-resolved BHXB radio/X-ray slopes. {{Source/campaign series with at least 10 detections and at least three measurements in each half are included. The suffix [MK] identifies a separately analysed MeerKAT/Swift campaign rather than an additional physical source.}} Positive $\Delta\xirx$ means that the radio/X-ray track steepens between the lower- and higher-$\lamx$ halves of the same source. {{$\lamx^{\rm low}$ and $\lamx^{\rm high}$ denote the median $\log_{10}(\LX/\LEdd)$ values of the lower and higher subsets, respectively; they characterize the sampled luminosity ranges and are not fitted transition luminosities.}}\label{tab:slopes}}
{
\begin{tabular*}{\textwidth}{@{\extracolsep{\fill}}llrrrrccc}
\toprule
Source & Track & $N_{\rm low}$ & $N_{\rm high}$ & $\lamx^{\rm low}$ & $\lamx^{\rm high}$ & $\xirx^{\rm low}$ & $\xirx^{\rm high}$ & $\Delta\xirx$ \\
\midrule
GX~339$-$4 & radio-loud & 36 & 39 & $-2.85$ & $-1.71$ & $0.62\pm0.06$ & $0.58\pm0.04$ & $-0.03\pm0.07$ \\
GX~339$-$4~[MK] & radio-loud & 19 & 19 & $-3.57$ & $-1.79$ & $0.49\pm0.06$ & $0.72\pm0.05$ & $+0.24\pm0.08$ \\
H~1743$-$322 & radio-quiet & 18 & 18 & $-3.12$ & $-1.78$ & $0.22\pm0.03$ & $1.53\pm0.06$ & $+1.31\pm0.07$ \\
MAXI~J1348$-$630 & radio-quiet & 19 & 20 & $-5.02$ & $-2.37$ & $0.30\pm0.02$ & $0.91\pm0.09$ & $+0.61\pm0.10$ \\
MAXI~J1820$+$070 & radio-loud & 8 & 8 & $-6.18$ & $-3.18$ & $0.77\pm0.18$ & $0.55\pm0.12$ & $-0.22\pm0.21$ \\
Swift~J1727.8$-$1613 & radio-quiet & 16 & 16 & $-4.53$ & $-0.73$ & $0.36\pm0.03$ & $0.99\pm0.03$ & $+0.63\pm0.04$ \\
Swift~J1753.5$-$0127 & radio-quiet & 34 & 35 & $-2.40$ & $-1.72$ & $0.40\pm0.04$ & $0.84\pm0.09$ & $+0.44\pm0.10$ \\
V~404~Cyg & radio-loud & 13 & 13 & $-5.38$ & $-2.74$ & $0.54\pm0.03$ & $0.62\pm0.06$ & $+0.08\pm0.07$ \\
XTE~J1118$+$480 & radio-loud & 10 & 11 & $-4.05$ & $-3.48$ & $0.68\pm0.09$ & $\cdots$ & $\cdots$ \\
XTE~J1752$-$223 & radio-quiet & 13 & 13 & $-2.76$ & $-1.93$ & $0.45\pm0.40$ & $0.72\pm0.35$ & $+0.27\pm0.53$ \\
\bottomrule
\end{tabular*}}

\vspace{2pt}
\begin{minipage}{0.98\textwidth}
\footnotesize\textit{Note.} The split is made at each source's median $\lamx$. XTE~J1118$+$480 has a narrow low-state luminosity baseline ($\Delta\lamx=0.57$~dex between the half-medians) and {{its higher-$\lamx$ slope is poorly constrained; the nominal bootstrap value}} $\xirx^{\rm high}=-0.01\pm0.42$ {{does not provide a useful estimate of the corresponding mass leverage and is therefore omitted from the summary columns and from}} Figure~\ref{fig:stateslopes} or Figure~\ref{fig:leverage}. {{The same median-split procedure is applied to MAXI~J1348$-$630, Swift~J1727.8$-$1613, MAXI~J1820$+$070, and the MeerKAT GX~339$-$4 campaign. The resulting quantities describe curvature across the observed luminosity range rather than fitted spectral-state transition points.}}
\end{minipage}

\end{table*}

\begin{table*}[tbp]
\centering\scriptsize
\caption{{{Systematic checks and limitations of the BHXB radio/X-ray diagnostic.}}\label{tab:methodrobust}}
\begin{tabular}{ll}
\toprule
Check & Result and limitation \\
\midrule
\parbox[t]{0.20\textwidth}{{{Monitoring coverage}}} &
\parbox[t]{0.71\textwidth}{{{The sample includes 155 detections from the Carotenuto, Hughes, Shaw, and MeerKAT/Swift monitoring data. MAXI~J1348$-$630 and Swift~J1727.8$-$1613 show positive median-split curvature, whereas MAXI~J1820$+$070 is consistent with no steepening.}}} \\[4pt]

\parbox[t]{0.20\textwidth}{{{Campaign dependence}}} &
\parbox[t]{0.71\textwidth}{{{The 38 retained MeerKAT GX~339$-$4 pairs give $\Delta\xirx=0.24\pm0.08$, showing that measurable curvature is not confined to historically radio-quiet tracks. Campaigns whose relative distance conventions cannot be verified are analysed separately.}}} \\[4pt]

\parbox[t]{0.20\textwidth}{{{Censoring and variability}}} &
\parbox[t]{0.71\textwidth}{{{Upper limits are retained in the data record but not fitted. The pair bootstrap does not model serial correlation, selection effects, interpolation covariance, or a full intrinsic-scatter distribution; its uncertainties therefore characterize the sampled detections rather than population-level significance.}}} \\[4pt]

\parbox[t]{0.20\textwidth}{{{Radio spectral conversion}}} &
\parbox[t]{0.71\textwidth}{{{Measured or published radio spectral indices are used for the dedicated campaigns. The MeerKAT/Swift pairs adopt $\alpha=0$; changing the common value to $\alpha=\pm0.3$ shifts $\log\LR$ by approximately $\pm0.18$ dex. Time-dependent spectral evolution can introduce additional slope changes.}}} \\[4pt]

\parbox[t]{0.20\textwidth}{{{Distance and mass scale}}} &
\parbox[t]{0.71\textwidth}{{{Published luminosity scales and explicitly stated mass assumptions are retained. Source-wise changes in distance or mass primarily translate the tracks in $\lamx$ and $\DeltaR$; their effect is quantified in the accompanying sensitivity analysis.}}} \\[4pt]
\bottomrule
\end{tabular}
\end{table*}

\begin{figure}[tbp]
\centering
\includegraphics[width=\columnwidth]{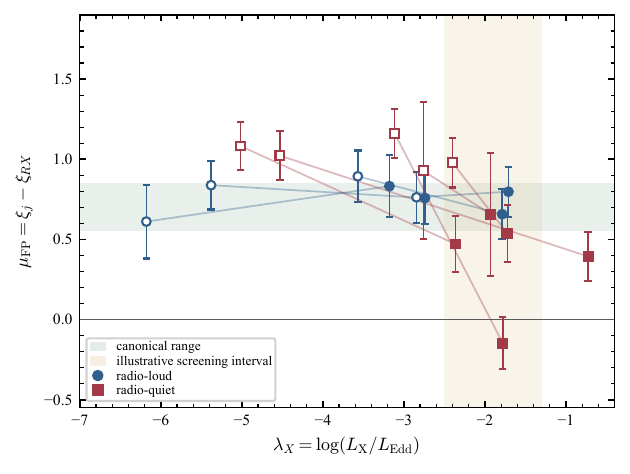}
\caption{Inferred inverse-FP mass leverage for the state-resolved BHXB slopes. Points show $\muFP=\xij-\xirx$ for the lower- and higher-$\lamx$ halves of each source, using $\xij=1.38\pm0.15$; the uncertainty includes the adopted uncertainty in $\xij$ and the fitted slope uncertainties. The green band marks the canonical FP mass-leverage range; {{the shaded vertical interval indicates the higher-luminosity range in which the empirical state diagnostic is most relevant and is not intended as a universal transition boundary.}} Open points denote lower-$\lamx$ halves and filled points denote higher-$\lamx$ halves. Values near or below zero indicate that a radio/X-ray inversion is no longer physically informative. XTE~J1118$+$480 is excluded (see Table~\ref{tab:slopes}).}
\label{fig:leverage}
\end{figure}

The implication for inverse-FP mass estimates is shown in Figure~\ref{fig:leverage}. We adopt a physically consistent hard-state jet exponent $\xij=1.38\pm0.15$. This follows from the identity $\muFP=\xij-\xirx$ using the canonical FP mass coefficient $\muFP=0.78$ and the reference hard-state slope $\xirx=0.60$, {{which gives $\xij=1.38$. The measured common slope of the historically radio-loud series would instead imply $\xij=1.355$, a difference well within the adopted $\pm0.15$ range.}} This value is close to the scale-invariant compact-jet prediction $17/12\simeq1.42$ from \citet{HeinzSunyaev2003}. To assess sensitivity, we bracket $\xij\in[1.23,1.53]$, corresponding to the adopted $\pm0.15$ uncertainty and including the theoretical value. The high-$\lamx$ slope of H~1743$-$322, $\xirx=1.53\pm0.06$, reaches the boundary $\muFP\simeq0$ even at the upper edge of this bracket, so the conclusion that the inversion becomes formally ill-conditioned is not a numerical artifact of choosing the canonical \citet{Merloni2003} ridge.

{{Taken together, the BHXB tracks show that radio/X-ray curvature is source dependent and can occur across a broad range of $\lamx$. The corresponding change in inverse-FP mass leverage follows only within the assumed jet/hot-flow scalings; at lower accretion rates, the applicability of those scalings must be established independently from the radio and X-ray emission properties.}}

\subsection{Additional monitoring and the scope of the state guide}\label{sec:additional}

{{The BHXB sample combines the archival database with several well-sampled monitoring campaigns that extend the luminosity and state coverage. For MAXI~J1348$-$630, we use the complete 44-row supplementary flux table of \citet{Carotenuto2021}, accessed through its machine-readable transcription in the Crook-Mansour repository and cross-matched to the original supplement for telescope and pairing information. Thirty-nine epochs have detections in both radio and X-rays and enter the diagnostic. Radio flux densities measured at 1.28~GHz (MeerKAT) or 5.5~GHz (ATCA) are converted to 5~GHz using $S_5=S_\nu(5/\nu)^\alpha$, adopting the tabulated spectral indices where available and the published mean value $\alpha=0.14$ otherwise. We adopt $D=2.2$~kpc and an assumed $M=10\,\Msun$. The published interpolated pairs and the flagged final extrapolated X-ray point are retained, while direct-pair and non-negative-spectral-index subsets are analysed separately as sensitivity checks.}}

{{For Swift~J1727.8$-$1613, we use the machine-readable, ejecta-corrected 5~GHz/1--10~keV luminosity table of \citet{Hughes2025}, corresponding to 32 detections at $D=2.6$~kpc; one radio upper limit is excluded from the fitted sample. The published spectral conversions and X-ray interpolation are retained, and an illustrative mass of $7\,\Msun$ is adopted for the state coordinate. For MAXI~J1820$+$070, all 16 radio/X-ray pairs listed by \citet{Shaw2021} are included at the published 0.1-dex luminosity precision, with $D=2.96$~kpc and the dynamical mass $M=8.48\,\Msun$. The original observing dates are retained, with a maximum radio/X-ray separation of 1.91~days.}}

{{The MeerKAT/Swift sample of \citet{CrookMansour2026} is taken from the per-source direct-pair tables. We retain BH or BH-candidate systems classified in the hard or quiescent state, require detections in both bands, and restrict the radio/X-ray separation to at most one day. GX~339$-$4 epochs identified as ejecta contaminated at MJD~58956--59080 and 59496--59505 are excluded. To avoid duplication with the dedicated MAXI~J1348$-$630, Swift~J1727.8$-$1613, and MAXI~J1820$+$070 campaigns, those sources are not reintroduced from the catalog sample. The resulting MeerKAT/Swift contribution contains 68 direct pairs at the catalog distances. Single-band 1.28-GHz luminosities are converted according to $L_R(5)=L_R(1.28)(5/1.28)$ for an adopted flat spectrum, $\alpha=0$; this conversion is a frequency normalization and does not by itself establish a flat compact-core spectrum.}}

{{For GX~339$-$4, H~1743$-$322, and IGR~J17091$-$3624, archival and MeerKAT/Swift measurements are treated as separate campaign series because their relative distance conventions cannot be verified from the inherited luminosity tables. The analysis therefore contains 24 source/campaign series representing 21 physical systems. Ten series satisfy the median-split sample-size requirement, and nine enter Figures~\ref{fig:stateslopes}--\ref{fig:leverage} after exclusion of the poorly constrained XTE~J1118$+$480 high-$\lamx$ fit. The [MK] suffix identifies a MeerKAT/Swift campaign; sources with sparse coverage or without a secure historical track classification remain in Figure~\ref{fig:xrbtracks} without being assigned to a radio-loud/radio-quiet class.}}

{{The resulting source-by-source behavior is not described by a single luminosity threshold. MAXI~J1348$-$630 has median-split slopes $0.30\pm0.02$ and $0.91\pm0.09$, while Swift~J1727.8$-$1613 gives $0.36\pm0.03$ and $0.99\pm0.03$. MAXI~J1820$+$070 shows no significant steepening, with $\Delta\xirx=-0.22\pm0.21$, whereas the independent MeerKAT GX~339$-$4 campaign gives $\Delta\xirx=+0.24\pm0.08$. These measurements therefore favor source-dependent curvature rather than a fixed division between standard and radio-quiet tracks.}}

{{The luminosity scales of the individual campaigns reinforce this interpretation. \citet{Carotenuto2021} locate the departure from the steep branch of MAXI~J1348$-$630 near $7\times10^{35}(D/2.2\,{\rm kpc})^2$~erg~s$^{-1}$ and its return toward the standard track near $10^{33}(D/2.2\,{\rm kpc})^2$~erg~s$^{-1}$. For the adopted mass, these correspond to $\lamx\simeq-3.26$ and $-6.10$. \citet{Hughes2025} obtain decay-only slopes of 0.32 and 1.42 for Swift~J1727.8$-$1613, with a break near $5.4\times10^{35}$~erg~s$^{-1}$, while inclusion of the rise changes the steeper branch to approximately 1.02. The locations and amplitudes of these track changes are therefore source and sampling dependent, rather than signatures of a universal hard-to-soft transition.}}

{{This behavior is also apparent below the illustrative $\lamx=-2.5$ reference scale. Twenty-six MAXI~J1348$-$630 and 21 Swift~J1727.8$-$1613 measurements lie below this value, with median reference-FP residuals of $-0.83$ and $-1.08$~dex, respectively, for the adopted masses and distances. By contrast, all 16 MAXI~J1820$+$070 pairs also lie below the same reference scale but have a median residual close to zero. Thus low $\lamx$ includes both FP-like and radio-deficient behavior. For MAXI~J1348$-$630, the whole-track bootstrap slope is $0.49\pm0.03$; restricting the analysis to nine directly paired epochs gives $0.47\pm0.07$, while excluding negative spectral indices gives $0.53\pm0.04$. These variations quantify the sensitivity of the whole-track slope to pairing and spectral selection.}}

\subsection{Adopted masses and inherited luminosity scale}\label{sec:sourceinputs}

{{Tables~\ref{tab:sourceinputs} and~\ref{tab:newinputs} summarize the mass and distance inputs used to place the BHXB measurements on the $\lamx$ axis. These quantities have heterogeneous levels of observational support: some masses are dynamical, whereas others are assumed, scaling based, or lower-limit proxies. They are therefore treated as state-coordinate inputs rather than as a uniformly precise dynamical sample. Logarithmic masses are rounded to two decimal places.}}

\begin{table*}[tbp]
\centering\scriptsize
\caption{{{Mass inputs for the 12 archival BHXBs. The luminosities are taken directly from the cited database release. Because the release luminosity table and its supplied derived products do not specify the source distance associated with each luminosity entry, no additional distance rescaling is applied here.}}\label{tab:sourceinputs}}
{
\begin{tabular}{lrrl}
\toprule
Source & $\mbh/\Msun$ & $\log(\mbh/\Msun)$ & Status of adopted mass \\
\midrule
A0620$-$00 & 6.6 & 0.82 & Dynamical (Cantrell et al. 2010) \\
GX~339$-$4 & 5.9 & 0.77 & Lower-limit proxy (Hynes et al. 2003) \\
H~1743$-$322 & 8.0 & 0.90 & Assumed \\
IGR~J17091$-$3624 & 10.0 & 1.00 & Assumed \\
MAXI~J1659$-$152 & 5.0 & 0.70 & Assumed \\
Swift~J1753.5$-$0127 & 7.4 & 0.87 & Assumed \\
V~404~Cyg & 9.0 & 0.95 & Dynamical (Khargharia et al. 2010) \\
XTE~J1118$+$480 & 7.6 & 0.88 & Dynamical (Gelino et al. 2006) \\
XTE~J1550$-$564 & 9.1 & 0.96 & Dynamical (Orosz et al. 2011) \\
XTE~J1650$-$500 & 5.0 & 0.70 & Uncertain; mass-function context (Orosz et al. 2004) \\
XTE~J1720$-$318 & 7.0 & 0.85 & Assumed \\
XTE~J1752$-$223 & 8.0 & 0.90 & Spectral-scaling proxy (Shaposhnikov et al. 2010) \\
\bottomrule
\end{tabular}}
\end{table*}

\begin{table*}[tbp]
\centering\scriptsize
\caption{{{Distance and mass inputs for the dedicated and MeerKAT/Swift source/campaign series. Distances are given in kpc and masses in solar masses. The catalog distance conventions are retained, while assumed or proxy masses are used only to place the measurements on the state coordinate. The [MK] suffix denotes a MeerKAT/Swift campaign.}}\label{tab:newinputs}}
{\begin{tabular}{lrrrl}
\toprule
Series & $N$ & $D$ & $M$ & Mass status \\
\midrule
MAXI~J1348$-$630 & 39 & 2.20 & 10.00 & assumed \\
Swift~J1727.8$-$1613 & 32 & 2.60 & 7.00 & assumed \\
MAXI~J1820$+$070 & 16 & 2.96 & 8.48 & dynamical \\
4U~1543$-$47~[MK] & 6 & 5.00 & 10.00 & assumed \\
EXO~1846$-$031~[MK] & 1 & 4.50 & 10.00 & assumed \\
GRS~1739$-$278~[MK] & 1 & 8.00 & 10.00 & assumed \\
GX~339$-$4~[MK] & 38 & 10.00 & 5.90 & inherited proxy \\
H~1743$-$322~[MK] & 4 & 8.50 & 8.00 & inherited proxy \\
IGR~J17091$-$3624~[MK] & 2 & 14.00 & 10.00 & inherited proxy \\
MAXI~J1631$-$479~[MK] & 8 & 5.10 & 10.00 & assumed \\
MAXI~J1803$-$298~[MK] & 2 & 8.00 & 10.00 & assumed \\
MAXI~J1810$-$222~[MK] & 6 & 8.00 & 10.00 & assumed \\
\bottomrule
\end{tabular}}
\end{table*}

{{A change in the adopted distance from $D$ to $D'$ and mass from $M$ to $M'$ shifts the state coordinate by
$\delta\lamx=2\log(D'/D)-\log(M'/M)$
and the reference residual by
$\delta\DeltaR=0.80\log(D'/D)-0.78\log(M'/M)$.
A common source-wise distance rescaling translates both luminosities and leaves the within-source radio/X-ray slope unchanged, although differences in distance convention between campaigns can introduce relative offsets. We therefore preserve the published luminosity scales rather than impose retroactive distance corrections that cannot be verified from the archival tables. The absolute $\lamx$ positions are correspondingly interpreted as state coordinates with known systematic uncertainty rather than as precision measurements of a universal transition luminosity.}}

\subsection{Sensitivity to the adopted FP calibration}\label{sec:fpcal}

The validity criterion is intentionally not tied to a single regression plane. For any inverse form
\begin{equation}
 \log\!\left(\frac{\mbh}{\Msun}\right)
 = \frac{\log\LR - \xirx\log\LX - b_0}{\muFP},
 \label{eq:inversefp}
\end{equation}
the calibration coefficients determine the absolute residual and the formal mass assigned to a source, but they do not determine whether the measured radio luminosity is a compact jet-base luminosity or whether the X-rays belong to the hard, radiatively inefficient branch. Table~\ref{tab:appcal} makes this distinction explicit. The numerical residuals for NGC~4395 move when the FP coefficients are changed, as expected, but the physical interpretation does not: the source remains compact-radio deficient relative to its reverberation mass. For the GC and $\omega$~Cen applications, the limiting issue occurs before a coefficient-dependent FP mass can be interpreted: the former lack independent accretor identity for the X-ray sources, and the latter lacks a detected compact radio core. Thus the classification of the examples is controlled by prerequisite failure modes, not by the choice of FP regression coefficients.

\begin{table*}[tbp]
\centering
\scriptsize
\setlength{\tabcolsep}{2.8pt}
\renewcommand{\arraystretch}{1.15}
\caption{Application-level robustness against representative FP calibrations. The table separates coefficient-dependent numerical residuals from coefficient-independent prerequisite failures.\label{tab:appcal}}
\begin{tabular*}{\textwidth}{@{\extracolsep{\fill}}lllll@{}}
\toprule
\parbox[t]{0.17\textwidth}{Application} & \parbox[t]{0.18\textwidth}{Merloni et al. (2003)} & \parbox[t]{0.18\textwidth}{G\"ultekin et al. (2009)} & \parbox[t]{0.18\textwidth}{Plotkin et al. (2012)} & \parbox[t]{0.22\textwidth}{Physical conclusion} \\
\midrule
\parbox[t]{0.17\textwidth}{NGC~4395, VLA scale} & \parbox[t]{0.18\textwidth}{{{$\DeltaR=-0.80$ dex; formal mass bias $\simeq-1.0$ dex}}} & \parbox[t]{0.18\textwidth}{$\DeltaR\simeq -1.1$ dex; low-bias sign unchanged} & \parbox[t]{0.18\textwidth}{$\DeltaR\simeq -0.3$ dex; low-bias sign unchanged} & \parbox[t]{0.22\textwidth}{Compact-core deficit; inverse FP would underestimate the reverberation mass if treated as a mass signal.} \\[4pt]
\parbox[t]{0.17\textwidth}{NGC~4395, VLBI jet base} & \parbox[t]{0.18\textwidth}{{{$\DeltaR<-2.16$ dex; formal mass bias $<-2.8$ dex}}} & \parbox[t]{0.18\textwidth}{$\DeltaR\lesssim -2.4$ dex} & \parbox[t]{0.18\textwidth}{$\DeltaR\lesssim -1.7$ dex} & \parbox[t]{0.22\textwidth}{The true jet-base constraint is more radio deficient than the VLA-scale measurement.} \\[4pt]
\parbox[t]{0.17\textwidth}{NGC~1399/NGC~4472 GC X-ray sources} & \parbox[t]{0.18\textwidth}{Coefficients are not yet physically interpretable.} & \parbox[t]{0.18\textwidth}{Same.} & \parbox[t]{0.18\textwidth}{Same.} & \parbox[t]{0.22\textwidth}{The limiting step is accretor identity; an FP mass would be a mass for an assumed compact object.} \\[4pt]
\parbox[t]{0.17\textwidth}{$\omega$~Centauri} & \parbox[t]{0.18\textwidth}{Radio limit constrains gas supply/core production.} & \parbox[t]{0.18\textwidth}{Same.} & \parbox[t]{0.18\textwidth}{Same.} & \parbox[t]{0.22\textwidth}{A radio non-detection is not a dynamical veto of the stellar-dynamical lower limit.} \\
\bottomrule
\end{tabular*}
\vspace{2pt}
\begin{minipage}{0.98\textwidth}
\footnotesize\textit{Note.} The NGC~4395 entries use the same published mass, X-ray luminosity, and radio measurements as Figure~\ref{fig:ngc4395}. The alternative-calibration entries are diagnostic translations intended to show the sign and approximate scale of the systematic response; they are not formal recalibrated inverse-FP masses. A formal comparison would require propagating each calibration-specific intercept, intrinsic scatter, sample selection, and X-ray band definition. The GC and $\omega$~Cen rows are intentionally not converted into coefficient-dependent masses because they fail prerequisite identity/core-detection checks before the FP can be physically inverted.
\end{minipage}
\end{table*}

\section{IMBH applications}\label{sec:application}

\subsection{NGC~4395}

\begin{figure}[tbp]
\centering
\includegraphics[width=\columnwidth]{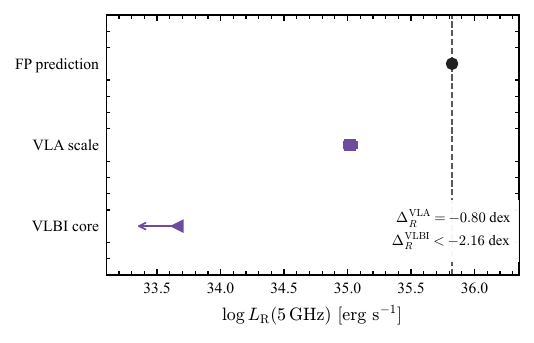}
\caption{NGC~4395 compact-core audit. The standard FP prediction, the VLA-scale luminosity, and the VLBI jet-base upper limit are shown on the same 5~GHz $\nu L_\nu$ axis. A radio deficit biases an inverse-FP mass low by $\sim|\DeltaR|/\muFP$: {{about 1.0~dex for the VLA-scale luminosity and at least 2.8~dex for the VLBI jet-base limit.}} This bias is in the opposite direction to the LMXB source-identity bias discussed for globular clusters, emphasizing that the failure mode depends on which prerequisite is violated. The VLA-scale point and VLBI jet-base limit are not interchangeable FP radio measurements: the former may include larger-scale nuclear emission, whereas the latter more directly constrains the unresolved jet-base component required by the FP.}
\label{fig:ngc4395}
\end{figure}

NGC~4395 is often treated as a low-mass AGN anchor because it has a reverberation mass $\log(\mbh/\Msun)=5.56\pm0.10$ \citep{Peterson2005} {{and a mean 2--10~keV luminosity $\log[\LX/(\mathrm{erg~s^{-1}})]=40.26$ ($\LX\simeq1.8\times10^{40}$~erg~s$^{-1}$) for $D=4.3$~Mpc \citep{King2013}. These values correspond to $\LX/\LEdd\simeq4.0\times10^{-4}$, or $\lamx\simeq-3.40$, for $\LEdd=1.26\times10^{38}(\mbh/\Msun)$~erg~s$^{-1}$. NGC~4395 therefore occupies a clearly low-Eddington regime, while its radio properties provide an independent test of whether the compact-core requirement is satisfied.}} It is valuable here precisely because it is not a clean anchor. The VLA-scale radio luminosity used by \citet{King2013} is already {{0.80~dex}} below the canonical FP prediction for the reverberation mass and mean X-ray luminosity. The deeper VLBI jet-base constraint is more than {{2.16~dex}} below the same reference value \citep{Wrobel2006,Yang2022}. Figure~\ref{fig:ngc4395} makes this a compact-core audit rather than a mass fit.

The radio deficit at the VLA scale translates to a downward mass bias of {{$\Delta\log\mbh\simeq\DeltaR^{\rm VLA}/\muFP\simeq-0.80/0.78\simeq-1.0$}}~dex if interpreted at face value. {{For the VLBI jet-base limit, the corresponding bias is $\Delta\log\mbh<-2.16/0.78\simeq-2.77$~dex, i.e. approximately 2.8~dex or more toward lower inferred mass.}} This illustrates that the bias direction for NGC~4395 is low, not high: the FP would underestimate the mass of a source with a genuine compact-core deficit. This is the opposite of the overestimate produced when a non-IMBH source, such as a globular-cluster LMXB, is incorrectly assigned an IMBH identity.

{These offsets are resolution-dependent compact-core diagnostics rather than independent precision inverse-FP mass estimates: the physical meaning of $\LR$ changes between the VLA-scale nuclear measurement and the VLBI jet-base limit.} The VLA, VLBI, and X-ray quantities are drawn from different observational epochs, and NGC~4395 is variable. Non-simultaneous radio and X-ray data can identify large compact-core deficits or contamination, but they should not be used for precision inverse-FP masses unless the variability amplitude is included or the observations are close enough in time to sample the same accretion state. A strictly quantitative inverse-FP test for this source would require simultaneous X-ray coverage and matched-resolution radio core measurements.

The lesson is general. A source can pass the low-Eddington test and still fail the compact-core test. For low-mass AGNs, the radio measurement must be resolved to, or at least consistent with, the unresolved jet base before the FP is inverted.

\subsection{Globular clusters}

\begin{figure}[tbp]
\centering
\includegraphics[width=\columnwidth]{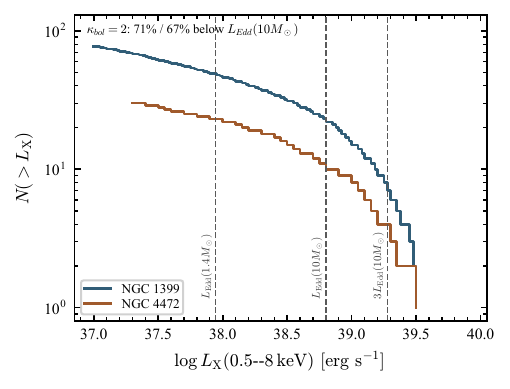}
\caption{Identity check for globular-cluster X-ray sources. The cumulative luminosity functions of NGC~1399 and NGC~4472 GC-associated X-ray sources are compared with illustrative Eddington reference lines for $1.4\,\Msun$, $10\,\Msun$, and $3\LEdd(10\,\Msun)$. The percentages mark the fractions below $\LEdd(10\,\Msun)$ for the adopted bolometric correction $\kappa_{\rm bol}=2.0$. The high-luminosity tail is an accretor-identity problem that precedes any inverse-FP mass measurement.}
\label{fig:gcs}
\end{figure}

In globular clusters, the limiting issue is different. The X-ray source may not be an IMBH at all. This point is especially important because radio/FP methods have long been proposed as efficient IMBH searches in GCs, and have been developed into modern ngVLA and SKA survey strategies \citep{Maccarone2004,MaccaroneFenderTzioumis2005,Wrobel2021,Karimi2024}. Figure~\ref{fig:gcs} shows the cumulative X-ray luminosity functions of GC-associated sources in NGC~1399 and NGC~4472.

\pauline{With the bolometric reference convention used in Figure~\ref{fig:gcs}, 71\% of the NGC~1399 sources and 67\% of the NGC~4472 sources lie below $\LEdd(10\Msun)$.} The remaining high-luminosity tail extends to the few-$\LEdd(10\Msun)$ regime, where bright neutron-star LMXBs, stellar-mass black-hole binaries, source blending, mild super-Eddington phases, or anisotropy can still be relevant. These luminosities therefore do not by themselves require IMBH accretors. The appropriate inference is not that globular clusters cannot host IMBHs, but that the FP test has not yet reached the mass-estimation stage for an individual GC X-ray source. It first fails at the identity-assignment stage: without an independently identified accreting black hole and compact radio core, an inverse-FP mass would be a mass for an assumed accretor, not a demonstrated IMBH. For GC sources, even a radio counterpart would not by itself be sufficient unless it is positionally coincident with the X-ray source or dynamical center, compact at the relevant angular scale, and inconsistent with a background AGN or diffuse cluster emission.

We note that the GC luminosities are measured in the 0.5--8~keV band \citep{Lehmer2020}, whereas the Eddington reference lines are computed from bolometric luminosities assuming a band correction factor of $\kappa=2.0$. Changing $\kappa$ shifts the visual placement of the Eddington reference lines, but it does not remove the main identity degeneracy: X-ray luminosity alone does not establish whether the accretor is an IMBH, a neutron star, or a stellar-mass black hole in a dense stellar environment.

This does not prove that globular clusters never host IMBHs. It shows that FP inversion is inappropriate for luminous cluster X-ray sources unless the accretor identity is independently established. Recent formation modelling also suggests that massive black holes in ordinary GCs should be rare, strengthening the need for identity checks before applying the FP \citep{Kritos2026}.

\subsection{$\omega$~Centauri}

The case of $\omega$~Centauri illustrates a distinct pitfall. Stellar-dynamical work has reported fast-moving central stars requiring a dark mass with a firm lower limit of $8.2\times10^3\,\Msun$ \citep{Haberle2024}. The subsequent ultradeep radio non-detection \citep{Mahida2026} should not be interpreted as a direct veto of the dynamical lower limit; it instead limits the observable accretion and compact-core output for any central black hole. {{\citet{Mahida2026} reach a similar physical conclusion, finding exceptionally inefficient accretion within their adopted model and deriving a conservative $3\sigma$ bound $\epsilon\lesssim4\times10^{-3}$. Here $\epsilon$ is treated as their model-dependent efficiency parameter rather than identified directly with $\LX/\LEdd$ or with a uniquely determined mass-inflow rate.}} A radio upper limit becomes a black-hole mass upper limit only after adopting a specific model for gas capture, accretion efficiency, jet production, and state. Without these assumptions, it is primarily an upper limit on observable compact-core output.

{{In this regime, the radio limit constrains the combined effects of gas supply, the fraction of captured material reaching the black hole, radiative efficiency, and compact-core production. Inferring $\dot m$ requires both the form and the normalization of the accretion-luminosity relation, so the scaling $\LX\propto\dot m^q$ alone does not define a unique accretion rate. The absence of a detected radio core therefore limits the observable accretion/jet output without, by itself, demonstrating jet absence or contradicting the dynamical mass constraint. This is a different failure mode from the loss of mass leverage associated with steep radio/X-ray tracks at higher accretion rates.}}


\subsection{Sgr~A*: a low-accretion comparison}\label{sec:sgra}

{{Sgr~A* provides a useful low-accretion comparison because its black-hole mass is securely established while its radiative output lies at an extremely low Eddington ratio. Its location relative to commonly used FP relations depends strongly on emission state and on the X-ray component being considered. In quiescence, Sgr~A* lies away from the standard FP relation, whereas bright X-ray flares move it closer to the expected locus \citep{Neilsen2013,Plotkin2012}. The contrast reflects the changing origin of the X-ray emission rather than a change in the black-hole mass: quiescent and flaring states probe different radiative processes and need not satisfy the same radio/X-ray scaling. Sgr~A* therefore illustrates that even for a source with a precisely known mass and very low $\lamx$, the applicability of the FP depends on whether the observed radio and X-ray components correspond to the physical regime represented by the calibration.}}

\section{Future observations}\label{sec:survey}

The current data do not yet provide the decisive IMBH-scale test. BHXBs give the state-resolved calibration; NGC~4395 gives a compact-core caution; globular-cluster X-ray populations give an identity warning; and $\omega$~Cen shows that a radio non-detection is not an exclusion. What is still missing is a sample of low-mass accretors with all four quantities measured simultaneously: an independent mass, a simultaneous X-ray luminosity and spectral state, a compact flat-spectrum radio core or a meaningful upper limit on the core, and enough repeat coverage to identify variability or state changes.

{{ESO~243$-$49 HLX-1 already satisfies several elements of this observational framework. X-ray disk modelling provides mass constraints that are independent of the FP, although they remain accretion-model dependent rather than dynamical \citep{Davis2011}. Repeated X-ray monitoring establishes clear state evolution, and radio emission has been detected both during transition-related flaring and during a hard-state interval \citep{Webb2012,Cseh2015}. The compact-core interpretation is less secure: the hard-state radio spectral index, $\alpha=-0.4\pm1.0$, is consistent with either flat or optically thin emission, the source remains unresolved on scales much larger than a jet base, and Doppler boosting may affect the observed radio luminosity \citep{Cseh2015}. HLX-1 is therefore a particularly useful intermediate case, with substantial state and temporal coverage but less secure constraints on the independent mass scale and on the nature of the radio core. Distinguishing transition flares from a persistent compact component remains essential before the source can serve as a controlled inverse-FP benchmark.}}

\begin{figure*}[tbp]
\centering
\includegraphics[width=0.76\textwidth]{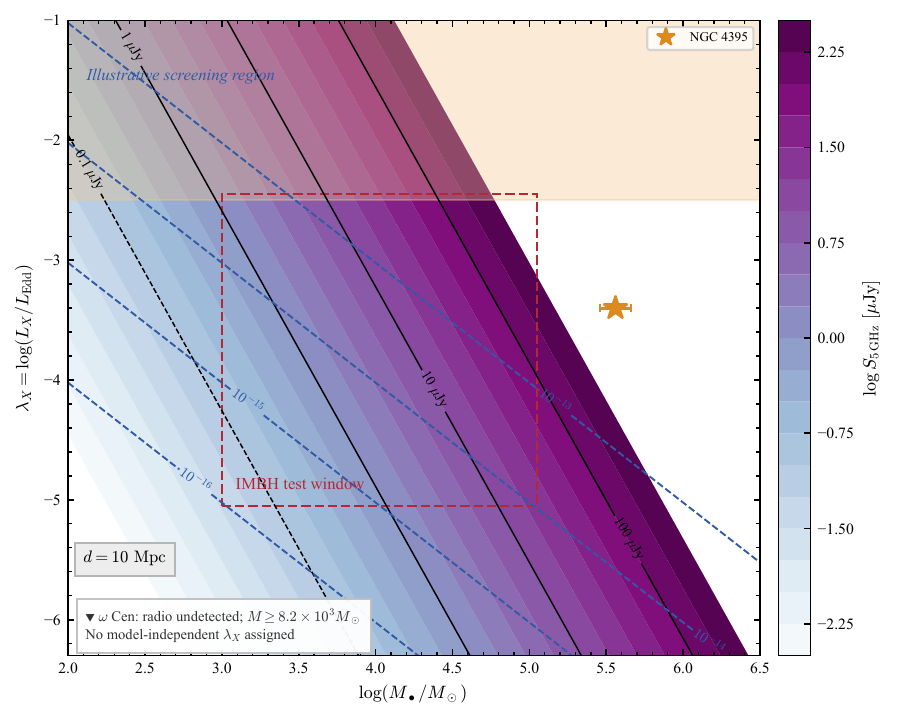}
\caption{Observation-design plane for a fiducial distance of 10~Mpc. Color shading shows the 5 GHz flux density predicted by the canonical FP; dashed contours show X-ray flux in erg~s$^{-1}$~cm$^{-2}$. The dashed red box marks the IMBH mass--state region where new observations would most directly test the proposed validity criterion. The radio contours assume the canonical FP coefficients, compact-core 5~GHz emission, and a flat radio spectrum unless otherwise stated. Fluxes scale as $d^{-2}$, so the 10~Mpc plane is an observing-design reference rather than a source-specific prediction. {{NGC~4395 is placed at its adopted mass and X-ray Eddington ratio. For $\omega$~Cen, only the radio non-detection and dynamical lower mass bound are indicated because its accretion luminosity cannot be assigned a model-independent $\lamx$. The shaded upper interval marks the higher-luminosity regime used as an observational reference in the BHXB analysis rather than a universal spectral-state boundary.}}}
\label{fig:survey}
\end{figure*}

Figure~\ref{fig:survey} recasts this missing data set as an observing problem. The red box can be interpreted directly as a compact-core sensitivity requirement. For the canonical FP at a fiducial distance of 10~Mpc, a $10^3\,\Msun$ black hole produces only $S_{5\,\mathrm{GHz}}\simeq0.04,0.14,0.57$, and $1.1\,\mu$Jy at $\lamx=-5,-4,-3$, and $-2.5$, respectively. A $10^4\,\Msun$ object gives $S_{5\,\mathrm{GHz}}\simeq0.86,3.4,13.6$, and $27\,\mu$Jy over the same state range, while a $10^5\,\Msun$ object reaches the tens-to-hundreds of $\mu$Jy regime (Table~\ref{tab:sensitivity}). Thus the low-mass, low-$\lamx$ part of the IMBH window is not simply a generic radio-detection problem; it is a sub-$\mu$Jy compact-core imaging problem. Current deep VLA observations can test the high-mass and/or high-$\lamx$ part of the window for the nearest systems, whereas the decisive $10^3$--$10^4\,\Msun$ regime requires ngVLA/SKA-class sensitivity and angular resolution. SKA helps most by converting this sensitivity into a candidate-selection engine: deep multi-frequency radio maps can identify sub-$\mu$Jy to few-$\mu$Jy compact-core candidates and prioritize those requiring matched X-ray/state confirmation.\par

A crucial point is that sensitivity alone is insufficient. At 10~Mpc, 1~pc subtends only $\simeq0.02^{\prime\prime}$ and 10~pc subtends $\simeq0.2^{\prime\prime}$. Therefore, an FP-usable radio detection must satisfy a joint sensitivity-resolution requirement: the emission must be localized to the candidate compact core and separated from diffuse star formation, supernova remnants, extended optically thin jet/lobe emission, and unrelated background sources. A nominal $\mu$Jy-level detection is not by itself a valid FP input unless its morphology, position, and spectrum are consistent with a compact jet base. \par

The most useful radio strategy is therefore multi-frequency compact-core imaging rather than a single-band detection. Matched-resolution observations at 5--10~GHz would test whether the candidate core has a flat or mildly inverted spectrum, as expected for a partially self-absorbed compact jet, or whether the radio emission is optically thin and likely contaminated by extended structure. ngVLA and SKA-Mid Band-5-like observations are naturally suited to this regime because they target the frequency range needed for compact-core identification and spectral-index measurements. A future SKA2 or ngVLA-depth program would push the same test toward lower masses, lower $\lamx$, and larger distances, where the expected FP core fluxes fall below the current few-$\mu$Jy regime. The physical role of SKA in this framework is therefore not simply to increase detection counts, but to separate FP-usable compact cores from contaminating radio components before any inverse-FP mass is quoted.

\begin{table}[tbp]
\centering
\scriptsize
\setlength{\tabcolsep}{3pt}
\renewcommand{\arraystretch}{1.12}
\caption{Canonical 5~GHz compact-core flux-density requirements at 10~Mpc for the Figure~\ref{fig:survey} IMBH test window. Values are in $\mu$Jy and assume the reference FP coefficients and a flat compact-core spectrum. Fluxes scale as $d^{-2}$.\label{tab:sensitivity}}
\begin{tabular*}{\columnwidth}{@{\extracolsep{\fill}}lrrrr@{}}
\toprule
$\mbh$ & $\lamx=-5$ & $\lamx=-4$ & $\lamx=-3$ & $\lamx=-2.5$ \\
\midrule
$10^3\,\Msun$ & 0.04 & 0.14 & 0.57 & 1.1 \\
$10^4\,\Msun$ & 0.86 & 3.4 & 13.6 & 27 \\
$10^5\,\Msun$ & 20.6 & 81.9 & 326 & 651 \\
\bottomrule
\end{tabular*}
\end{table}

{{A decisive test will require low-mass accretors with independent mass constraints and repeated, component-resolved radio/X-ray measurements across changes in accretion state. Such a sample would allow FP residuals to be tracked together with radio spectral shape, X-ray emission mechanism, and compact-core strength, directly testing whether departures from the adopted relation follow changes in the underlying accretion flow. The observational requirement is therefore not simply deeper radio detections, but a state-resolved and spatially resolved radio/X-ray data set in which simultaneity, source identification, and angular resolution are integral to the mass diagnostic.}}

\section{Conclusions}\label{sec:conclusions}

We have formulated a physical validity test for inverse-FP IMBH masses. The main conclusions are:

\begin{enumerate}

\item In the scale-invariant compact-jet/hot-flow limit, the inverse-FP mass leverage is $\muFP=\xij(1-1/q)=\xij-\xirx$. As the radio/X-ray track steepens or the compact jet is suppressed, a mass inversion becomes ill-conditioned rather than merely less precise. The identity $\muFP=\xij-\xirx$ holds because $\xirx=\xij/q$; both forms are equivalent within the ADAF/compact-jet model.

\item {{The 424 radio/X-ray detections from 21 BHXBs and candidates reveal substantial source-dependent diversity in track shape and curvature.}} {{Historically standard tracks can remain close to $\xirx\simeq0.6$, whereas several radio-quiet or hybrid systems show pronounced slope evolution over their sampled luminosity ranges;}} H~1743$-$322 reaches $\xirx=1.53\pm0.06$ in its high-$\lamx$ half. {{The independent MeerKAT/Swift GX~339$-$4 campaign also shows measurable curvature, demonstrating that such behavior is not restricted to historically radio-quiet systems.}} The track classification follows \citet{Gallo2012} and \citet{Coriat2011} independently of the steepening test.

\item NGC~4395 is not a clean FP anchor. It has a low Eddington ratio but a compact-radio deficit of {{$-0.80$}}~dex on VLA scales and {{$<-2.16$}}~dex at the VLBI jet-base level, {{corresponding to formal downward mass biases of about 1.0~dex and at least 2.8~dex, respectively, if the radio deficits are interpreted as mass signals.}} This is a low bias, opposite in sign to the LMXB misidentification bias.

\item Globular-cluster X-ray populations in NGC~1399 and NGC~4472 are source-identity limited. With the bolometric convention adopted here, 71\% and 67\% of the associated sources lie below $\LEdd(10\Msun)$, respectively, while the remaining high-luminosity tail does not by itself demonstrate an IMBH accretor. For these sources the limiting issue is source identity, not FP calibration.

\item A future FP-based IMBH mass should be reported only after four checks: independent source identity, {{an accretion and emission state independently shown to be compatible with the adopted FP relation}}, a compact, positionally coincident radio core whose spectrum is flat or mildly inverted, whose emission is not dominated by star formation, supernova-remnant emission, extended lobes, or a background source, and whose measurement is simultaneous or near-simultaneous with the X-ray state measurement where variability is important, and a small FP residual (applicable only when $\mbh$ is independently known). {{The X-ray Eddington ratio can help identify the relevant regime, but the BHXB results show that no single value of $\lamx$ is sufficient to establish FP applicability.}} For SKA-era searches, this means that radio detections and upper limits should first be filtered by compactness, spectral index, positional coincidence, and simultaneity before being converted into FP masses. The missing high-value observation is a simultaneous X-ray/state and high-resolution radio-core campaign for low-mass AGNs and dynamically selected IMBH candidates in the $10^3$--$10^5\,\Msun$ range.

\end{enumerate}

The Fundamental Plane remains a valuable tool, but its inverse use should be conditional. \pauline{{{Tables~\ref{tab:methodrobust} and~\ref{tab:appcal} summarize the observational systematics and calibration dependence that accompany this physical applicability test.}}} The main contribution of this paper is not a new universal coefficient set, but a physically transparent rule for deciding when the FP is allowed to constrain an IMBH mass at all.

\section*{Acknowledgments}
This work used the radio/X-ray binary database of \citet{Bahramian2023} and NASA's Astrophysics Data System. We thank Pauline Barmby for valuable suggestions that improved the observational framing and clarity of the manuscript. The analysis scripts and derived data tables used to generate the figures will be made available with the published article or upon reasonable request.

\section*{Data Availability}
{{The BHXB analysis uses archival luminosities from \citet{Bahramian2023} together with the radio/X-ray monitoring data of \citet{Carotenuto2021,Hughes2025,Shaw2021,CrookMansour2026}. The reproducibility package contains the public input files and source manifest, the normalized detection table, the observation-selection ledger, adopted source parameters and sensitivity tables, the executed analysis notebook, and the figure products. Where the same system appears in both a dedicated monitoring study and the broader MeerKAT/Swift catalog, the dedicated campaign is used to avoid duplicate epochs. Campaigns whose relative distance conventions cannot be verified from the archival luminosity products are analysed separately. The archival database does not provide sufficient information to reconstruct all source distances and observing dates associated with the published luminosities, and these remain provenance limitations. The NGC~4395 and globular-cluster data are retained as independent application samples and do not enter the BHXB radio/X-ray fits.}}

\bibliographystyle{aasjournal}
\bibliography{references}

@misc{Bahramian2023,
 author = {{Bahramian}, A. and {Rushton}, A.},
 title = {{Radio/X-ray correlation database for X-ray binaries, v220908}},
 year = {2022}, publisher = {Zenodo}, doi = {10.5281/zenodo.7059313}
}

@article{Bariuan2022,
  author  = {{Bariuan}, L. G. C. and {Snios}, B. and {Sobolewska}, M. and {Siemiginowska}, A. and {Schwartz}, D. A.},
  year    = {2022},
  journal = {\mnras},
  volume  = {513},
  pages   = {4673}
}

@incollection{Belloni2010,
  author    = {{Belloni}, T. M.},
  title     = {{States and Transitions in Black Hole Binaries}},
  booktitle = {{The Jet Paradigm}},
  editor    = {{Belloni}, T. M.},
  series    = {Lecture Notes in Physics},
  volume    = {794},
  pages     = {53},
  publisher = {Springer},
  address   = {Berlin},
  year      = {2010}
}

@article{Corbel2013,
  author  = {{Corbel}, S. and {Coriat}, M. and {Brocksopp}, C. and others},
  year    = {2013},
  journal = {\mnras},
  volume  = {428},
  pages   = {2500}
}

@article{Coriat2011,
  author  = {{Coriat}, M. and {Corbel}, S. and {Prat}, L. and others},
  year    = {2011},
  journal = {\mnras},
  volume  = {414},
  pages   = {677}
}

@article{Falcke2004,
  author  = {{Falcke}, H. and {K\"ording}, E. and {Markoff}, S.},
  year    = {2004},
  journal = {\aap},
  volume  = {414},
  pages   = {895}
}

@article{FenderBelloniGallo2004,
  author  = {{Fender}, R. P. and {Belloni}, T. M. and {Gallo}, E.},
  year    = {2004},
  journal = {\mnras},
  volume  = {355},
  pages   = {1105}
}

@article{Gallo2012,
  author  = {{Gallo}, E. and {Miller}, B. P. and {Fender}, R.},
  year    = {2012},
  journal = {\mnras},
  volume  = {423},
  pages   = {590}
}

@article{Gultekin2009,
  author  = {{G\"ultekin}, K. and {Cackett}, E. M. and {Miller}, J. M. and others},
  year    = {2009},
  journal = {\apj},
  volume  = {706},
  pages   = {404}
}

@article{Gultekin2019,
  author  = {{G\"ultekin}, K. and {King}, A. L. and {Cackett}, E. M. and others},
  year    = {2019},
  journal = {\apj},
  volume  = {871},
  eid     = {80},
  pages   = {80}
}

@article{Gultekin2022,
  author  = {{G\"ultekin}, K. and {Nyland}, K. and {Gray}, N. and others},
  year    = {2022},
  journal = {\mnras},
  volume  = {516},
  pages   = {6123}
}

@article{Haberle2024,
  author  = {{H\"aberle}, M. and {Neumayer}, N. and {Seth}, A. and others},
  year    = {2024},
  journal = {\nat},
  volume  = {631},
  pages   = {285}
}

@article{HeinzSunyaev2003,
  author  = {{Heinz}, S. and {Sunyaev}, R. A.},
  year    = {2003},
  journal = {\mnras},
  volume  = {343},
  pages   = {L59}
}

@article{Karimi2024,
  author  = {{Karimi}, B. and {Barmby}, P. and {Abbassi}, S.},
  year    = {2024},
  journal = {\apj},
  volume  = {974},
  eid     = {260},
  pages   = {260},
  doi     = {10.3847/1538-4357/ad77c9}
}

@article{King2013,
  author  = {{King}, A. L. and {Miller}, J. M. and {G\"ultekin}, K. and others},
  year    = {2013},
  journal = {\apj},
  volume  = {771},
  eid     = {84},
  pages   = {84}
}

@misc{Kritos2026,
  author        = {{Kritos}, K. and {Wadekar}, D. and {Berti}, E.},
  title         = {{Predicting intermediate-mass black hole formation in star clusters with machine learning}},
  year          = {2026},
  archivePrefix = {arXiv},
  eprint        = {2605.21593},
  primaryClass  = {astro-ph.GA}
}

@article{Lehmer2020,
  author  = {{Lehmer}, B. D. and {Eufrasio}, R. T. and {Tzanavaris}, P. and others},
  year    = {2020},
  journal = {\apjs},
  volume  = {248},
  eid     = {31},
  pages   = {31}
}

@article{Maccarone2004,
  author  = {{Maccarone}, T. J.},
  year    = {2004},
  journal = {\mnras},
  volume  = {351},
  pages   = {1049}
}

@article{MaccaroneFenderTzioumis2005,
  author  = {{Maccarone}, T. J. and {Fender}, R. P. and {Tzioumis}, A. K.},
  year    = {2005},
  journal = {Ap\&SS},
  volume  = {300},
  pages   = {239},
  doi     = {10.1007/s10509-005-1188-5}
}

@article{Maccarone2003,
  author  = {{Maccarone}, T. J. and {Kundu}, A. and {Zepf}, S. E.},
  year    = {2003},
  journal = {\apj},
  volume  = {586},
  pages   = {814}
}

@article{Mahida2026,
  author  = {{Mahida}, A. D. and {Bahramian}, A. and {Miller-Jones}, J. C. A. and {Sett}, S. and {Dage}, K. and {Strader}, J. and {Galvin}, T. J. and {Paduano}, A.},
  title   = {{No Evidence for Accretion around the Intermediate-mass Black Hole in Omega Centauri}},
  year    = {2026},
  journal = {\apj},
  volume  = {996},
  eid     = {122},
  pages   = {122},
  archivePrefix = {arXiv},
  eprint  = {2512.09649},
  primaryClass = {astro-ph.HE}
}

@article{Merloni2003,
  author  = {{Merloni}, A. and {Heinz}, S. and {di Matteo}, T.},
  year    = {2003},
  journal = {\mnras},
  volume  = {345},
  pages   = {1057}
}

@article{Peterson2005,
  author  = {{Peterson}, B. M. and {Bentz}, M. C. and {Desroches}, L.-B. and others},
  year    = {2005},
  journal = {\apj},
  volume  = {632},
  pages   = {799}
}

@article{Plotkin2012,
  author  = {{Plotkin}, R. M. and {Markoff}, S. and {Kelly}, B. C. and {K\"ording}, E. and {Anderson}, S. F.},
  year    = {2012},
  journal = {\mnras},
  volume  = {419},
  pages   = {267}
}

@article{Wang2024,
  author  = {{Wang}, Y. and {Wang}, T. and {Ho}, L. C. and {Zhong}, Y. and {Luo}, B.},
  year    = {2024},
  journal = {\aap},
  volume  = {689},
  eid     = {A327},
  pages   = {A327}
}

@article{Wrobel2006,
  author  = {{Wrobel}, J. M. and {Ho}, L. C.},
  year    = {2006},
  journal = {\apjl},
  volume  = {646},
  pages   = {L95}
}

@article{Wrobel2021,
  author  = {{Wrobel}, J. M. and {Maccarone}, T. J. and {Miller-Jones}, J. C. A. and {Nyland}, K. E.},
  year    = {2021},
  journal = {\apj},
  volume  = {918},
  eid     = {18},
  pages   = {18},
  doi     = {10.3847/1538-4357/ac0ef3}
}

@article{Yang2022,
  author  = {{Yang}, X. and {Yao}, S. and {Yang}, J. and others},
  year    = {2022},
  journal = {\mnras},
  volume  = {514},
  pages   = {6215}
}

@article{YuanNarayan2014,
  author  = {{Yuan}, F. and {Narayan}, R.},
  year    = {2014},
  journal = {\araa},
  volume  = {52},
  pages   = {529}
}

@article{Carotenuto2021,
 author={{Carotenuto}, F. and {Corbel}, S. and {Tremou}, E. and others},
 title={{The hybrid radio/X-ray correlation of the black hole transient MAXI J1348-630}},
 year={2021}, journal={\mnras}, volume={505}, pages={L58--L63}, doi={10.1093/mnrasl/slab049}
}

@article{Hughes2025,
 author={{Hughes}, A. K. and {Carotenuto}, F. and {Russell}, T. D. and others},
 title={{The peculiar hard state behaviour of the black hole X-ray binary Swift J1727.8-1613}},
 year={2025}, journal={\mnras}, volume={542}, pages={1803--1816}, doi={10.1093/mnras/staf1363}
}

@article{Shaw2021,
 author={{Shaw}, A. W. and {Plotkin}, R. M. and {Miller-Jones}, J. C. A. and others},
 title={{Observations of the Disk/Jet Coupling of MAXI J1820+070 during Its Descent to Quiescence}},
 year={2021}, journal={\apj}, volume={907}, pages={34}, eprint={2012.04024}, archivePrefix={arXiv}
}

@article{CrookMansour2026,
 author={{Crook-Mansour}, J. and {Fender}, R. and {Hughes}, A. and others},
 title={{The homogeneous MeerKAT and Swift/XRT X-ray binary radio:X-ray plane}},
 year={2026}, journal={\mnras}, volume={551}, pages={stag1022}, doi={10.1093/mnras/stag1022}
}

@article{Cseh2015,
 author={{Cseh}, D. and {Webb}, N. A. and {Godet}, O. and others},
 title={{On the radio properties of the intermediate-mass black hole candidate ESO 243-49 HLX-1}},
 year={2015}, journal={\mnras}, volume={446}, pages={3268--3276}, doi={10.1093/mnras/stu2363}
}

@article{Webb2012,
 author={{Webb}, N. and {Cseh}, D. and {Lenc}, E. and others},
 title={{Radio Detections During Two State Transitions of the Intermediate-Mass Black Hole HLX-1}},
 year={2012}, journal={Science}, volume={337}, pages={554--556}, doi={10.1126/science.1222779}
}

@article{Davis2011,
 author={{Davis}, S. W. and {Narayan}, R. and {Zhu}, Y. and others},
 title={{The Cool Accretion Disk in ESO 243-49 HLX-1: Further Evidence of an Intermediate-Mass Black Hole}},
 year={2011}, journal={\apj}, volume={734}, pages={111}, eprint={1104.2614}, archivePrefix={arXiv}
}

@article{Neilsen2013,
 author={{Neilsen}, J. and {Nowak}, M. A. and {Gammie}, C. and others},
 title={{A Chandra/HETGS Census of X-Ray Variability from Sgr A* during 2012}},
 year={2013}, journal={\apj}, volume={774}, pages={42}, doi={10.1088/0004-637X/774/1/42}
}

\end{document}